\documentclass{nature}

\title{Tunneling, Remanence, and Frustration in Dysprosium based Endohedral Single Molecule Magnets}

\author{R. Westerstr\"om$^{1,2,3}$, J. Dreiser$^{2}$, C. Piamonteze$^{2}$, M.  Muntwiler$^{2}$, S. Weyeneth$^1$, K. Kr\"amer$^{4}$,\\ S.X. Liu$^{4}$, S. Decurtins$^{4}$, A. Popov$^{5}$, S. Yang$^{5,6}$, L. Dunsch$^{5}$, \& T. Greber$^1$}

\begin{document}

\maketitle

\begin{affiliations}
\item 	Physik-Institut, Universit\"at Z\"urich,  Switzerland
\item 	Swiss Light Source, Paul Scherrer Institut,  Switzerland
\item	Department of Physics and Astronomy, Uppsala University,  Sweden
\item	Departement f\"ur Chemie und Biochemie, Universit\"at Bern, Switzerland
\item	Department of Electrochemistry and Conducting Polymers,Leibniz Institute of Solid 		State and Materials Research, Germany
\item	Hefei National Laboratory for Physical Sciences at Microscale,
		Department of Materials Science and Engineering, University of Science and 					Technology of China, China
\end{affiliations}


\begin{abstract}
A single molecule magnet (SMM) can maintain its magnetization direction over a long period of time \cite{sessoliNat93,GatteschiBook06}. It consists in a low number of atoms that facilitates the understanding and control of the ground state, which is essential in future applications such as high-density information storage or quantum computers \cite{leuenbergerNat01,bogani08mNat}. Endohedral fullerenes realize robust, nanometer sized, and chemically protected magnetic clusters that are not found as free species in nature. Here we demonstrate how adding one, two, or three dysprosium atoms to the carbon cage results in three distinct magnetic ground states. The significantly different hysteresis curves demonstrate the decisive influence of the number of magnetic moments and their interactions. At zero field the comparison relates tunneling of the magnetization, with remanence, and frustration. The ground state of the tridysprosium species turns out to be one of the simplest realizations of a frustrated, {\it{ferromagnetically}} coupled magnetic system.
\end{abstract}

The discovery of single molecule magnets containing one single lanthanide ion triggered large interest in 4$f$ electron compounds \cite{Ishikawa03,wodruffChemRev13}. 
However, the remarkable double decker molecules with one magnetic 4$f$ ion have poor remanence:
The zero field magnetization decays rapidly, also via the unavoidable tunneling between states with opposite magnetization.  In this respect, dinuclear $4f$ compounds appear to be more robust due to exchange coupling related stabilization of the magnetic moments, \cite{Guo11,RinehartJACS2011,RinehartNatChem11,mondal12,habibChemSocRev13}
and there are reports on trinuclear lanthanide ion complexes with \cite{tangACI2006} and without \cite{Luzon2008,LiviuAngew08} magnetic ground states.

Endohedral fullerenes \cite{pop13} represent a new family in the class of lanthanide-based single molecule magnets. 
They can contain clusters that bear great potential when it comes to the production of molecular arrays on surfaces.
Many of them are particularly stable, survive sublimation and may be easily imaged  \cite{treierPRB,zhao11} and manipulated with scanning probes \cite{butcher03,CPL12}. 
While the R=holmium or terbium based R$_3$N@C$_{80}$ species showed non-collinear paramagnetism \cite{wolfJMag05}, it was recently found that the isotropic gadolinium R$_3$ species, shows ferromagnetically coupled collinear paramagnetic behaviour \cite{Nafradi2012}.  
However, the first endofullerene which displayed hysteresis and qualified as a single molecule magnet was DySc$_{2}$N@C$_{80}$ \textbf{1}. 
The observed hysteresis is a result of a slow relaxation of the magnetization which is caused by a ligand field that splits the Hund ground state and causes barriers separating states with different magnetization  \cite{westerstromJACS}. Here we present results for the complete dysprosium-scandium endofullerene series\\ Dy$_{n}$Sc$_{3-n}$N@C$_{80}$ ($n=\textbf{1},\textbf{2},\textbf{3}$) with one, two, or three  4$f$ moments inside a nanometer sized closed shell C$_{80}$ cage (Fig. 1a). 
This bottom up approach of building a magnet features the unique opportunity to study the effect of adding moments - one by one. 
In zero field the magnetization of $n=\mathbf{1}$ decays via quantum tunneling, while ferromagnetic coupling of the individual dysprosium moments results in remanence for $n=\mathbf{2}$ and in a frustrated ground state for $n=\mathbf{3}$.

Dy$_{n}$Sc$_{3-n}$N@C$_{80}$ ($n=\textbf{1},\textbf{2},\textbf{3}$) (isomer I$_h$), hereafter the isomeric label is omitted for clarity, were produced by a modified Kr\"atschmer-Huffman dc-arc discharge method in a mixture of NH$_3$ (20 mbar) and He (200 mbar) atmosphere \cite{dunsch04,yangPhysChemB07,yangS09}. To ensure a low background signal for the SQUID measurements the molecules were drop cast onto  sample holders with a weak linear diamagnetic behavior made from kapton foil. This diamagnetic background has been subtracted from the data. 
For GdCl$_3$~6H$_2$O (Aldrich) our magnetometer shows at 6~K from the observed Brillouin function a Gd magnetic moment of 7.4$\pm$0.2 $\mu_B$, which compares to 7$\mu_B$ as expected from the Gd$^{3+}$ $^8$S$_{7/2}$ ground state.
To obtain the relaxation times at elevated temperatures, the ac susceptibility of \textbf{2} was measured for varying frequencies of the oscillating driving magnetic field. 

In zero field the interaction between the magnetic moments of $n$ magnetic atoms may be described with a Hamiltonian reminiscent to Heisenberg and Lines \cite{lines,luz12} of the form:
\begin{equation}
\mathcal{H}=\sum_{i\neq k}^{n} j_{i,k } {~}{\textbf{J}}_i\cdot {\textbf{J}}_k
\label{eq1}
\end{equation}
where $j_{i,k}$ are the coupling constants and ${\textbf{J}}_{i,k}$ the corresponding angular momentum operators on sites $i$ and $k$, respectively.
For Ho and Tb trimetal nitride endofullerenes it was proposed that the magnetic moments $\mu_i$, which are parallel to the expectation values $\langle \textbf{J}_i \rangle$, remain aligned in the R$^{3+}$ - N$^{3-}$ ligand field \cite{wolfJMag05}.
Our findings on \textbf{1} \cite{westerstromJACS} and ab-initio results \cite{Liviu} are in line with the picture where the $\langle \textbf{J}_i \rangle$ of every Dy$^{3+}$ are uniaxial (anisotropic). 
This allows the reduction of the ground state problem to a non-collinear Ising model with $n$ pseudospins \cite{LiviuAngew08}, which can take two orientations, parallel or antiparallel to the corresponding Dy-N axis.
The 2$^n$ solutions for such a Hamiltonian form 2$^{n-1}$ doublets.
They are labeled TRD since the two states are time reversal symmetric and have opposite magnetization but the same energy in zero field (see Fig. 1c).
Importantly, the interaction $j_{i.k}$ between the different pseudospins lifts the degeneracy of the 2$^{n-1}$ TRD's and gives rise to excitation energies U$_n$.
For \textbf{1} the solution is trivial since no interaction occurs. 
The tunneling rate between the two states in the single doublet determines the magnetization time.
For \textbf{2} the two TR doublets are split by the interaction $j_{1,2}$.
This causes remanence, because demagnetization involves the excitation into the second TRD, or instantaneous tunneling of the two magnetic moments.
With the same $j_{i,k}$ between all ions in \textbf{3}, which is given if the ions sit on an equilateral triangle, we find the four TRD's to split in a group of three magnetic and one non-magnetic doublet. 
The fact that \textbf{3} shows paramagnetic behaviour indicates a negative $j_{i,k}$, i.e. ferromagnetic coupling.
This imposes for  \textbf{3} a six fold degenerate ground state, where tunneling between these six states enables demagnetization.
The appearance of three TR doublets of anisotropic, ferromagnetically coupled pseudospins results in magnetic frustration.
Notably, this is analogous to the case of isotropic spins on an equilateral triangle, where frustration is caused by anti-ferromagnetic exchange interaction \cite{seo13}. 

The pseudospin structures of the ground states for  \textbf{1-3} are shown in Fig. 1c.
The level scheme of the 2$^{n-1}$ TR doublets is reflected in the magnetization curves.
The magnetic moment of a given molecule corresponds to the vectorial sum of the $n$ individual moments.
In a magnetic field the TRD's undergo Zeeman splitting, and since they are different for \textbf{1-3}, distinct susceptibility, beyond scaling with $n$ is observed.
In Fig. 1d the magnetizations at the temperature of 6 K are displayed as a function of the applied field.
The curves for the three molecules are different, not only due to the number of Dy atoms per molecule, as can be seen in Fig. 1e. The relative differences between the three molecules amount up to 10\%, which allows the extraction of the different ground state parameters.

The magnetization curves are reminiscent to Brillouin functions, though, in the present case the Dy$^{3+}$ moments do not align along the magnetic field and the degeneracy of the $^6$H$_{15/2}$ ground state is partly lifted by the ligand field.
Assuming randomly frozen, independent molecules, reduces the saturation magnetization to half the value of the maximum magnetization of free molecules, since only the projection on the field direction contributes.
Together, with the given structure of the TRD's we can extract the average magnetic moments $\mu_{i,n}$ and the TRD splittings U$_2$ and U$_3$ (Fig. 1c) from comparison of simulated magnetization curves with the experiment.
The solid lines in Fig. 1d represent the best fits of these simulations to the measured data.
The average magnetic moments $\mu_{i,n}$ and the excitation energies U$_n$ of \textbf{1-3}, are listed in Table \ref{T1}.

The magnetic moment of \textbf{1}  of 9.37~$\mu_B$ agrees with a large $m_J$ ground state along the Dy-N axis.
For \textbf{2}  the splitting U$_2$ between the two TRD's is in the order of 1~meV.
For \textbf{3} the value of U$_3$ indicates a weaker coupling than in \textbf{2}.

Fig. 2a  displays magnetization curves at 2~K taken with a field sweep rate of 0.8 mTs$^{-1}$ for \textbf{1}-\textbf{3}. 
The observed hystereses demonstrate that the rate at which the magnetization relaxes to its equilibrium is slow compared to the measurement time, which is characteristic for single molecule magnets.
The distinct shapes indicate on how strong the number of magnetic moments and their interaction influence the response to external magnetic field changes.
For applications the remanence, i.e. the memory of magnetization history in zero-field is of particular interest. 
There is large "remanence"  for \textbf{2}, as compared to a sharp drop of the magnetization at low fields for~\textbf{1}, and a narrow hysteresis with vanishing zero-field magnetization for \textbf{3}.
It is a clear consequence of the magnetic interaction between the endohedral dysprosium ions in \textbf{2} and \textbf{3}, which is mediated by the central N$^{3-}$ ion.
For \textbf{1} the enhanced tunneling of magnetization in the absence of an applied field is seen in the abrupt jump of the magnetization when approaching the {$\mu_0H~=~0$} point.
The narrow hysteresis of \textbf{3} makes it the softest single molecule magnet of the three. This is due to magnetic frustration of the ground state, which suppresses remanence.
The Zeeman splitting between the lowest and the first excited state in \textbf{3} is smaller than in \textbf{1}, which allows more efficient flipping of the magnetization, also in an applied field.
So far frustration was not realized in trinuclear magnetic molecules as the relevant mechanism for zero field demagnetization \cite{wolfJMag05,Luzon2008,LiviuAngew08}.
In contrast to \textbf{1} and \textbf{3} the reversal of magnetization in \textbf{2} requires a {\it{simultaneous}} flip of both magnetic moments or the crossing of the barrier U$_2$, which consequently stabilizes the zero field magnetization. The barrier has contributions from the exchange energy and the dipolar coupling of the individual moments $\mu_{i,2}$.

In Fig. 2b the Hilbert space of the time reversal doublets ($n,\pm d$) for the three molecules (see Fig. 1c) are shown. 
$\pm d$ are the indices of the two states in the given TRD's. The 2$^n$ states are connected by a network of single tunneling transitions that correspond to the flipping of one magnetic moment. For \textbf{1} and \textbf{3}  all ground state TRD's are connected by single-tunneling transitions at the ground state energy, which is an intrinsic demagnetization mechanism that suppresses remanence.
For \textbf{2} there is no single-tunneling path connecting the ground state TRD, and a single-tunneling event costs the energy U$_2$.
   
The U$_2$ barrier is also reflected in the temperature dependence of the zero field magnetization decay times.
Below 5 K a double exponential was fitted to the decay data (Fig. 3a), as done for
 the case of DySc$_2$N@C$_{80}$, where this behaviour was ascribed to different hyperfine interaction of different Dy isotopes \cite{westerstromJACS}.
The resulting decay times for the slower process, $\tau_A$, are displayed on a logarithmic scale versus the reciprocal temperature in Fig. 3b. 
A 100 s blocking temperature of about 5.5 K is determined, which is amongst the highest temperatures reported for single molecule magnets \cite{RinehartNatChem11,RinehartJACS2011}.
Higher temperatures were accessed using ac magnetic susceptibility measurements and the corresponding relaxation times are displayed as open symbols in Fig. 3b.
Clearly, the relaxation times show two temperature regimes, indicating distinct relaxation mechanisms.
Down to 2~K the zero-field relaxation times do not show a temperature independent region, as observed for a single pseudospin flip tunneling regime in \textbf{1} \cite{westerstromJACS}, because this relaxation mechanism is suppressed in the ground state of \textbf{2} by the barrier U$_2$.

Fitting the lifetimes $\tau_A$ to:
\begin{equation}
\tau_A=\frac{\tau_1\cdot\tau_2}{\tau_1+\tau_2}
\label{eq2}
\end{equation} 
leads with $\tau_{\ell} = \tau_{2,\ell}^0\exp(U_{2,\ell}^{\mathrm{eff}}/k_BT)$ to the solid curve in Fig. 3b.
The effective energy barriers for magnetization reversal get $U_{2,1}^{\mathrm{eff}} = 0.73 \pm 0.04$ meV and $U_{2,2}^{\mathrm{eff}} = 4.3 \pm 0.1$ meV with pre-exponential factors $\tau_{2,1}^0= 56.5 \pm 9.8$ s and $\tau_{2,2}^0 = 12.0 \pm 1.3$ ms, respectively.
The lower barrier  $U_{2,1}^{\mathrm{eff}}$ corresponds to the energy gap between the two TR doublets of \textbf{2} (Fig. 1c and Table \ref{T1}).
The higher temperature barrier $U_{2,2}^{\mathrm{eff}}$  must be related to relaxation via higher lying excited states.
The value for $U_{2,2}^{\mathrm{eff}}$ is similar to the one found in a Co$_2$Dy$_2$ compound \cite{mondal12}.
As in \textbf{1} \cite{westerstromJACS} the prefactors $\tau^{0}_{2,\ell}$ in \textbf{2} are, compared to other Dy based single molecule magnets \cite{Ishikawa03,mondal12}, remarkably large.
This is taken as an indication that the phase spaces for tunneling and excitations leading to decay of the magnetization are particularly small, which must be due to the peculiar protection of the magnetic moments in the closed shell C$_{80}$ cage.

In summary, the three dysprosium based endofullerenes Dy$_{n}$Sc$_{3-n}$N@C$_{80}$ ($n=1,2,3$) are identified as single molecule magnets with three different ground states. 
The present pseudospin model for the ground states is expected to be generally valid for all uniaxially anisotropic R$_3$N@C$_{80}$ endofullerenes.
The distinct hysteresis curves reflect on how dramatic changes can be caused by stoichiometry and interaction in single molecule magnets.
The observed large remanence in \textbf{2} is due to an energy barrier for flips of individual 4$f$ moments.
For the trinuclear nitrogen-cluster Dy$_{3}$N@C$_{80}$ the ferromagnetic coupling results in a frustrated ground state that suppresses remanence regardless of the exchange and dipolar barrier. 
These findings demonstrate the crucial role of magnetic frustration for the suppression of magnetization blocking in single molecule magnets.

\begin{figure}
\caption{\textbf{Ground state magnetic structure}. \textbf{a}, Ball- and stick-model of R$_3$N@C$_{80}$ R = Rare earth (here Dy or Sc). \textbf{b}, Model of the endohedral R$^{3+}_3$N$^{3-}$ unit and the corresponding couplings $j_{i,k}$ that are partly mediated across the N$^{3-}$ ion. \textbf{c}, Ground state magnetic structure for Dy$_n$Sc$_{3-n}$N@C$_{80}$ based on 2$^{n-1}$ ferromagnetically coupled time reversal symmetric doublets ($n,\pm d$) for $n=\textbf{1-3}$, where $d$ is the doublet index. The energies U$_2$ and U$_3$ are the exchange and dipole barriers for \textbf{2} and \textbf{3} respectively.  \textbf{d}, Magnetisation m$(\mu_0H)$ of \textbf{1-3} at 6 K. The experimental data (dots) are scaled to the magnetic moment per molecule as obtained from the fits of the three ground states in \textbf{c}. \textbf{e}, Deviation of m/m$_{sat}$ of  \textbf{2} and \textbf{3} from \textbf{1}.}
\end{figure}

\begin{figure}
\caption{\textbf{Magnetic hysteresis loops.} \textbf{a}, Hysteresis curves for \textbf{1}-\textbf{3} recorded using SQUID magnetometry at 2~K at a field sweep rate of 0.8 mTs$^{-1}$. 
\textbf{b}, Hilbert space topology of the 2$^n$ pseudospin states ($n,\pm d$) in \textbf{1}-\textbf{3}. Solid lines correspond to single tunneling events of one magnetic moment between two states at the same energy. Red dashed lines involve an energy barrier, which is due to exchange and dipolar coupling.}
\end{figure} 

\begin{figure}
\caption{\textbf{Magnetic zero field relaxation times}. \textbf{a}, Zero-field relaxation curves for \textbf{2} after saturation at $\mu_0$H = 7~T. $m_{sat}$ is the magnetization at 7 T. The line corresponds to a fit of a double- ($T<4.5$ K) and a single- ($T>4.5$ K) exponential. \textbf{b}, Corresponding relaxation times $\tau_A$ as a function of inverse temperature. Open symbols are ac susceptibility results. The red line is the fit to Eq. (\ref{eq2}).}
\end{figure}


\begin{addendum}
 \item[Acknowledgement] The project is supported by the Swiss National Science
		Foundation (SNF project 200021 129861 and 147143), the Swedish research council 			(350-2012-295) and the Deutsche Forschungsgemeinschaft (DFG project PO 1602/1-1). 
		We gratefully acknowledge fruitful discussions with Liviu Chibotaru. Version submitted for peer review on August 23 2013. 
 \item[Competing Interests] The authors declare that they have no
		competing financial interests.
 \item[Correspondence] Correspondence and requests for materials
should be addressed to Thomas Greber ~(email: greber@physik.uzh.ch).
  \end{addendum}

\begin{table}
\caption{\textbf{Magnetic moments and lowest excitation energies.} Parameters from the fit of the magnetization curves to the level scheme in Fig. 1c.
The average magnetic moments per Dy ion $\mu_{i,n}$ are given in $\mu_B$, the excitation energies U$_n$ in meV. U$^{\mathrm{eff}}_{2,1}$ is the excitation energy extracted from the zero field relaxation times (see Fig. 3b).}
\label{T1}
\begin{center}
\begin{tabular}{|l|c|c|c|}
\hline
& $\mu_{i,n}$ &U$_{n}$ &U$^{\mathrm{eff}}_{n,1}$ \\
\hline
DySc$_2$N@C$_{80}$ & 9.37$\pm$ 0.06 &-&- \\
\hline
Dy$_2$ScN@C$_{80}$ & 8.75$\pm$ 0.13&0.96$\pm$0.1&0.73$\pm$0.04 \\
\hline
Dy$_3$N@C$_{80}$& 9.46$\pm$ 0.05&0.30$\pm$0.2&-\\
\hline
\end{tabular}
\end{center}
\end{table}

\end{document}